\documentclass[aps,prl,twocolumn,showpacs,superscriptaddress,amsmath,amssymb,longbibliography]{revtex4-1}

\usepackage{amsmath}
\usepackage{graphicx}
\usepackage{float}
\usepackage{color}

\begin{document}

\title{Ordering of Binary Colloidal Crystals by Random Potentials}

\author{Andr\'e S. Nunes}
\email{contributed equally.}
\affiliation{Centro de F\'{\i}sica Te\'orica e Computacional and Departamento de F\'{\i}sica, Faculdade
	de Ci\^{e}ncias, Universidade de Lisboa, P-1749-016 Lisboa, Portugal, EU}
	
\author{Sabareesh K. P. Velu}
\email{contributed equally.}
\affiliation{Department of Physics, Bilkent University, Cankaya, 06800 Ankara, Turkey}

\author{Iryna Kasianiuk}
\affiliation{Department of Physics, Bilkent University and UNAM, Cankaya, 06800 Ankara, Turkey}

\author{Denys Kasyanyuk}
\affiliation{Department of Physics, Bilkent University and UNAM, Cankaya, 06800 Ankara, Turkey}

\author{Agnese Callegari}
\affiliation{Department of Physics, Bilkent University and UNAM, Cankaya, 06800 Ankara, Turkey}

\author{Giorgio Volpe}
\affiliation{Department of Chemistry, University College London, 20 Gordon Street, London WC1H 0AJ, United Kingdom, EU}

\author{Margarida M. Telo da Gama}
\affiliation{Centro de F\'{\i}sica Te\'orica e Computacional and Departamento de F\'{\i}sica, Faculdade
	de Ci\^{e}ncias, Universidade de Lisboa, P-1749-016 Lisboa, Portugal, EU}

\author{Giovanni Volpe}
\affiliation{Department of Physics, Bilkent University, Cankaya, 06800 Ankara, Turkey}
\affiliation{Department of Physics, University of Gothenburg, 41296 Gothenburg, Sweden, EU}

\author{Nuno A. M. Ara\'ujo}
\email{nmaraujo@fc.ul.pt}
\affiliation{Centro de F\'{\i}sica Te\'orica e Computacional and Departamento de F\'{\i}sica, Faculdade
	de Ci\^{e}ncias, Universidade de Lisboa, P-1749-016 Lisboa, Portugal, EU}

\begin{abstract}
Structural defects are ubiquitous in condensed matter, and not always a nuisance. For example, they underlie phenomena such as Anderson localization and hyperuniformity, and they are now being exploited to engineer novel materials. Here, we show experimentally that the density of structural defects in a 2D binary colloidal crystal can be engineered with a random potential. We generate the random potential using an optical speckle pattern, whose induced forces act strongly on one species of particles (strong particles) and weakly on the other (weak particles). Thus, the strong particles are more attracted to the randomly distributed local minima of the optical potential, leaving a trail of defects in the crystalline structure of the colloidal crystal. While, as expected, the crystalline ordering initially decreases with increasing fraction of strong particles, the crystalline order is surprisingly recovered for sufficiently large fractions. We confirm our experimental results with particle-based simulations, which permit us to elucidate how this non-monotonic behavior results from	 the competition between the particle-potential and particle-particle interactions. 
\end{abstract}

\maketitle

\begin{figure}[b]
\includegraphics[width=\columnwidth]{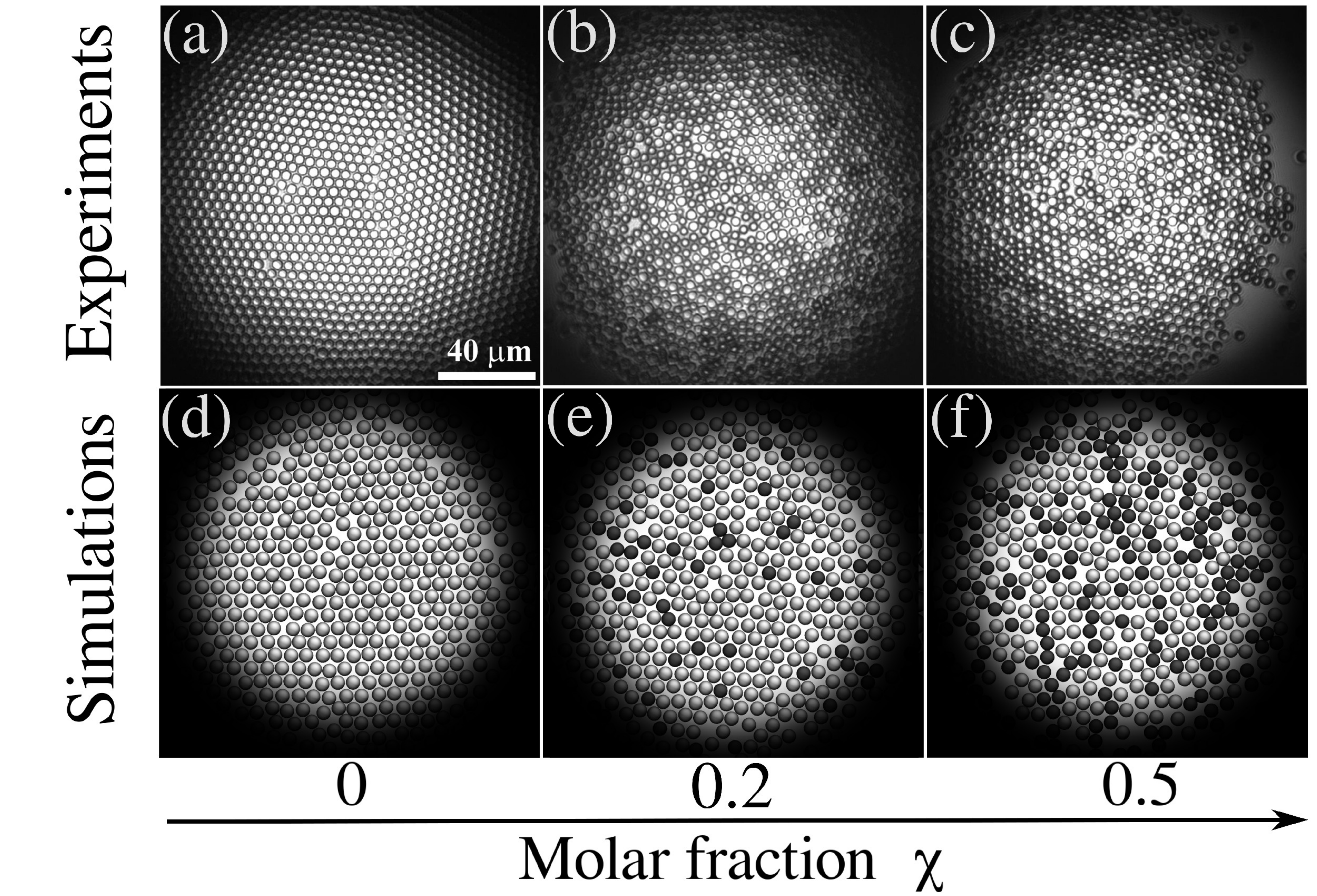} 
\caption{\textbf{Colloidal crystals with tunable degree of disorder.}
Final configurations obtained in (a-c) experiments and (d-f) simulations, for different molar fractions $\chi$ of strong particles. The weak (silica) particles are light gray, and the strong (polystyrene) particles are dark gray.
}
\label{fig:1}
\end{figure}

Perfect crystalline structures are not commonly found in Nature, because, even in the absence of impurities, structural defects occur spontaneously and disrupt the periodicity of the crystalline lattice \cite{Bollmann2012}. For example, when a melt is cooled down, multiple crystallites grow with degenerate orientations \cite{Pyka2013}. 
Since the coarsening time of these crystallites diverges with size, structural defects appear and prevent the emergence of global order \cite{Zurek2009, Campo2010}.
While the existence of these defects is a challenge when growing single crystals, it can also be an opportunity when  engineering the properties of materials; indeed, control over defects enables the development of solid-state devices with  fine-tuned mechanical resilience, optical properties, and heat and electrical conductivity \cite{Lifshitz1966, Hurle2004, Chen2016, Boeck2017, Heyde2013}.
In atomic crystals, engineering structural defects is an experimental challenge for two reasons \cite{Kulkarni2005}: 
first, current visualization techniques at the atomic scale do not provide a high spatial or time resolution \cite{Kramer2010, Faleev2013}; second, no current technique can control the density of defects in a systematic manner \cite{Wang2018}. 
The first challenge can be overcome studying colloidal crystals as models for atomic systems \cite{Deutschlander2015, Irvine2012}, where colloidal particles can be individually tracked using standard digital video microscopy techniques  \cite{Nunes2016, Pham2016, Brazda2017}. 
Here, we demonstrate that the second challenge can be solved combining a binary colloidal mixture and an optical random potential generated by a speckle light pattern. This permits us to control the density of structural defects in the resulting 2D colloidal crystal and to explore a surprising non-monotonic behavior of their ordering and stability.

We use a binary colloidal suspension of equally-sized polystyrene (refractive index $n_{\rm ps}\approx1.59$) and silica ($n_{\rm si}\approx1.42$) spherical particles with diameters $d_{\rm PS} = 4.06 \pm  0.11\,{\rm\mu m}$ and $d_{\rm SiO_2} = 3.93 \pm  0.12\,{\rm\mu m}$, respectively. 
To characterize the composition of the mixture, we use the molar fraction of polystyrene particles defined as $\chi = N_{ \rm ps}/N_{\rm t}$ where $N_{\rm ps}$ is the number of polystyrene particles and $N_{\rm t}$ is the total number of particles.
We let these particles sediment at the bottom surface of a homemade sample chamber so that they are effectively confined in a quasi-2D space (see Supplemental Material for details~\cite{SM}).
We illuminate from above with a speckle pattern, which we generate by mode-mixing a coherent laser beam in a multimode optical fibre (see supplementary Fig.~S1 and Supplemental Material for details~\cite{SM}) \cite{Volpe2014, Volpe2014OE, Pince2016}. Speckle patterns form rough, disordered optical potentials characterized by wells whose depths are exponentially distributed and whose average width is given by diffraction (here, average grain size $\sigma = 3.75 \pm 0.2\ \mu{\rm m}$). 
Furthermore, the fibre imposes a Gaussian envelope (beam waist $\sigma_{\rm G} = 72.5 \pm 0.2\ \mu{\rm m}$) to the speckle pattern, which attracts the particles towards the center of the speckle pattern effectively confining them in space.
Since the optical forces acting on the particles increase for larger mismatches between their refractive index and that of the surrounding medium (here water, $n_{\rm w}\approx1.33$) \cite{Jones2015}, the optical forces acting on the polystyrene (strong) particles are about $2\times$ higher than those exerted on silica (weak) particles. Importantly, the optical forces at the deepest local minima of the speckle potential are strong enough to trap the strong particles, but not the weak ones.

\begin{figure}[ht!]
\includegraphics[width=\columnwidth]{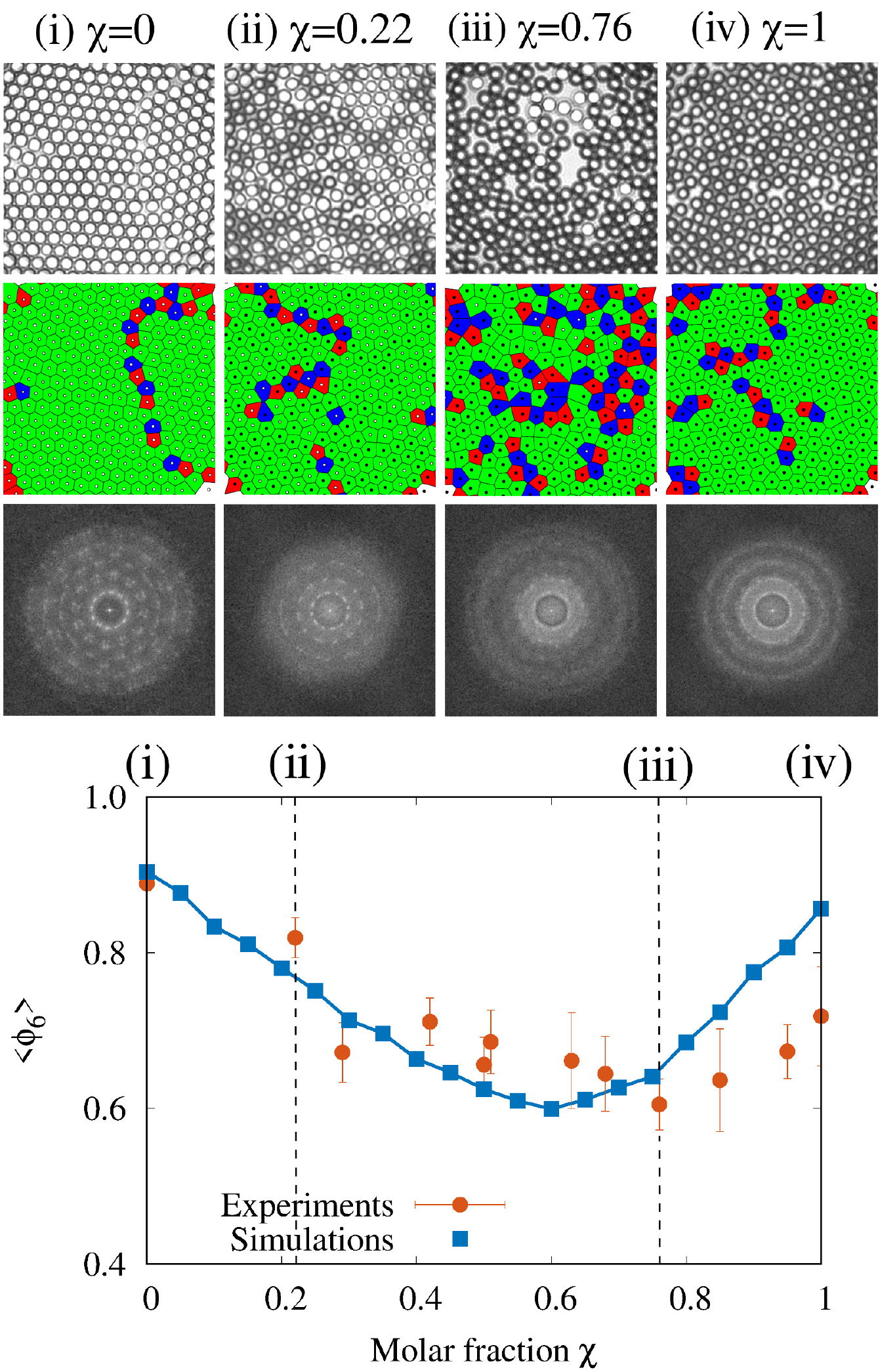} 
\caption{\textbf{Crystalline order for different molar fractions of strong particles.}
Six-fold bond order parameter $\langle\phi_6\rangle $ as a function of the molar fraction $\chi$ obtained experimentally (triangles) and numerically (squares; the blue line connects the symbols for visual guidance). The numerical results are averages over $100$ samples. 
The top snapshots show the final configurations in the experiments (first row), the Voronoi tesselation (second row), and the spatial Fourier transform (third row) for $\chi=0$, $0.22$, $0.76$, and $1$. The filled (empty) circles at the center of the Voronoi cells indicate strong (weak) particles. The cells are colored by the number of nearest neighbors, namely, equal (green), lower (red), greater (blue) than six.
See also supplementary video 1. 
}
\label{fig:2} 
\end{figure}

We start with a low concentration of particles ($1.4 \cdot 10^7\,{\rm mL^{-1}}$) and switch on the optical potential. The particles are attracted towards its center by the Gaussian envelope. When only weak particles  are present ($\chi=0$), they eventually form an (almost) perfect hexagonal colloidal crystal, as shown in Fig.~\ref{fig:1}a. When we introduce strong particles as $\chi$ increases, these get trapped in the local minima of the disordered potential and introduce defects that reduce the hexagonal order. Already with only $20\%$ of strong particles ($\chi=0.2$), the presence of structural defects is clearly visible (see Fig.~\ref{fig:1}b). The impact is even more pronounced when $50\%$ of the particles ($\chi=0.5$) are strongly interacting with the potential (Fig.~\ref{fig:1}c). 
Thus, strong particles act as defects in the crystalline structure of the weak ones, compromising global order. These results are confirmed by particle-based simulations, as shown in Figs.~\ref{fig:1}d-f (see supplemental information [19]). As we will see in more detail below, we can control the density of defects by adjusting $\chi$ as well as the intensity and grain size of the pattern.

To quantify the order of the crystalline structure, we measured the six-fold bond-order parameter, $\langle\phi_6\rangle$, defined as \cite{Nunes2018}
\begin{equation}\label{Eq:bond}
\langle\phi_6\rangle 
= 
\frac{1}{6N_c}\sum_{l}^{N_{\rm c}}
\left| 
\sum_{j}^{N_{\rm b}}e^{i6\theta_{lj}}
\right| ,
\end{equation}
where the outer sum is over the $N_{\rm c}$ particles within 7.5 particle diameters from the center of the potential (which is the area where the aggregate is formed), the inner sum is over the $N_{\rm b}$ neighbors of a particle in the Voronoi tessellation, and $\theta_{lj}$ is the angle between the $x$-axis and the line connecting the centers of particles $j$ and $l$. 
$\langle\phi_6\rangle=1$ for perfect hexagonal crystals (in practice, it is never exactly one, because of thermal fluctuations and other transient perturbations to the periodic order) and it decreases with the number of structural defects. 
Figure~\ref{fig:2} shows $\langle\phi_6\rangle$ obtained experimentally and numerically as a function of the molar fraction $\chi$. 
For $\chi=0$, $\langle\phi_6\rangle \approx 1$, consistent with the formation of an hexagonal periodic structure. 
As expected, as $\chi$ increases, the value of $\langle\phi_6\rangle$ decreases due to the formation of structural defects. 
The snapshots in the top rows of Fig.~\ref{fig:2} show the final configurations (first row), the corresponding Voronoi tessellations (second row), and the spatial Fourier transform (third row), for different values of $\chi$. 

\begin{figure*}[ht!]
\includegraphics[width=\textwidth]{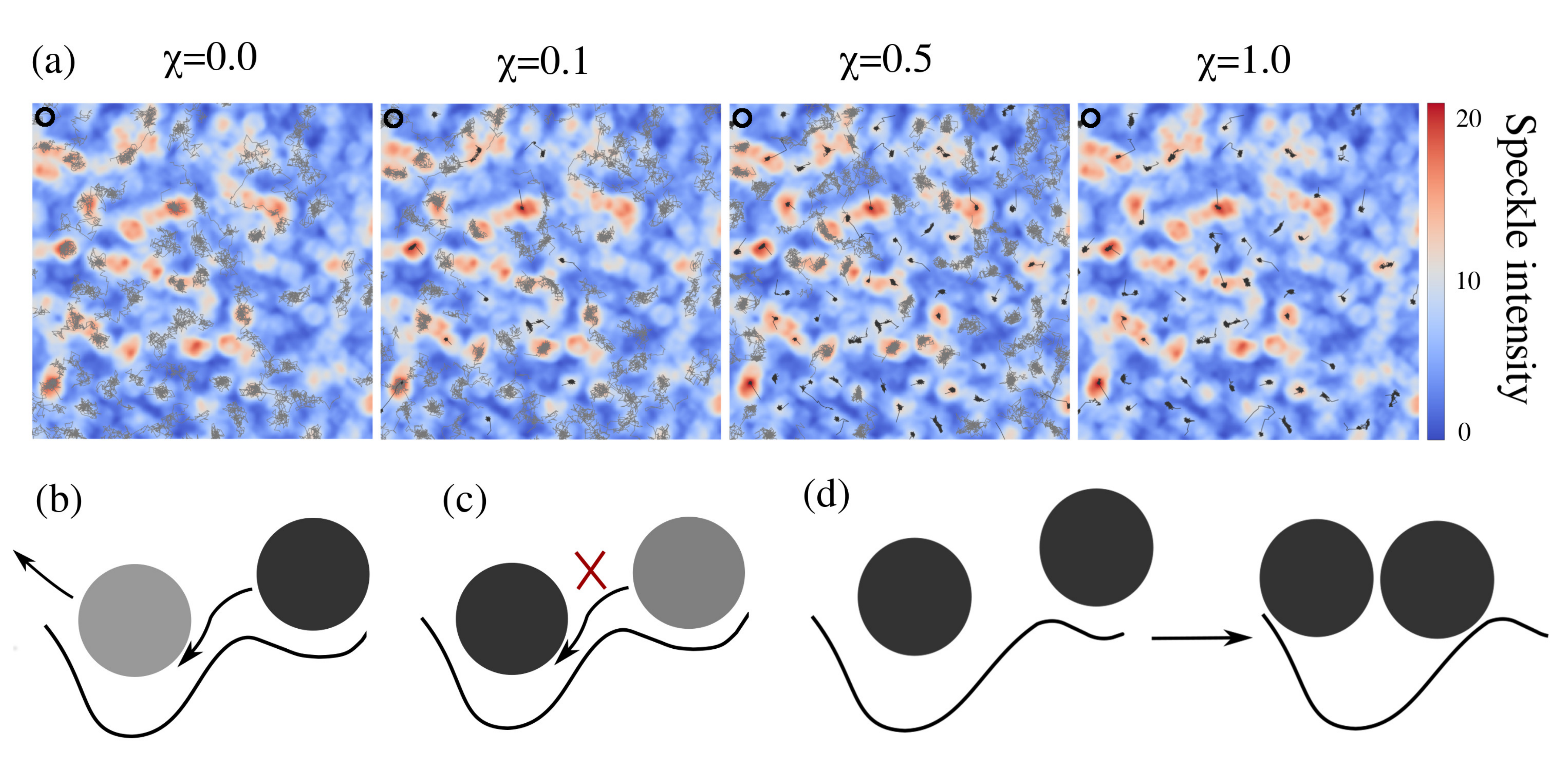}
\caption{\textbf{Local dynamics of the interaction between particles and minima in the random potential.}
(a) Examples of trajectories of weak (light gray) and strong (dark gray) particles in the presence of a speckle obtained numerically for different values of the molar fraction $\chi$. The particle density is $10\times$ lower than that of maximal packing and the Gaussian envelope is absent. The four simulations were preformed under exactly the same conditions, including the same sequence of random numbers for the thermostat (see Supplemental Material~\cite{SM}). The black circles on the top left corner indicate the particle size. The random potential intensities are in units of $k_{\rm B} T$ and $\sigma$ is one particle diameter.
(b) When a weak particle (light gray) is located at a potential minimum and a strong particle (dark gray) is in its vicinity, it is energetically favorable to exchange the two, but the opposite process (c) is not. 
(d) The free energy may be significantly reduced when two particles of the same species share the same potential minimum.
See also supplementary video 2.
}
\label{fig:3}
\end{figure*}

Surprisingly, the data reported in Fig.~\ref{fig:2} show that $\langle\phi_6\rangle$ reaches its minimum value for $\chi_{\rm min}\approx0.6$, and then the global order increases for $\chi>\chi_{\rm min}$. In particular, for $\chi=1$, the strong particles self-assemble into an hexagonal crystal, despite the presence of the underlying random potential. This result is corroborated by the Voronoi tessellation of the final configurations and by the respective spatial Fourier transforms. From this analysis, we can see that the number of Voronoi cells with a number of neighbors different from six becomes higher near the minimum of $\langle\phi_6\rangle$, and that the Fourier transforms display dimmer intensity peaks near the same value. This observation suggests a change in the effective interaction between the strong particles and the underlying potential: from one that favors disorder at a low $\chi$ to one favoring order at larger $\chi$. 

In order to elucidate the microscopic mechanisms underlying this behavior, we employ trajectories obtained by particle-based simulations to study the interactions between the two particle's species and the local minima in the potential.
Figure~\ref{fig:3}(a) shows some trajectories of weak (light gray) and strong (dark gray) particles at various $\chi$. 
We performed these simulations using a random potential without the Gaussian envelope to highlight the dynamics of the interaction between the particle and the local minima.
In all cases, the weak particles can hop between minima, while the strong particles get readily trapped in them; in fact, the effective diffusion coefficient of the strong particles is significantly lower than that of the weak particles (see supplementary Fig.~S2 \cite{SM}).
At low $\chi$, the strong particles quickly populate the minima that are sufficiently deep to prevent their escape and remain there for the entire simulation time, because this configuration is energetically favorable (Figs.~\ref{fig:3}b and \ref{fig:3}c); therefore, the number of spatial defects increases monotonically with the number of the trapped strong particles, leading to a decrease of $\langle\phi_6\rangle$ with increasing $\chi$. 
At large $\chi$, the number of strong particles is greater than the potential minima and thus it becomes energetically favorable to have more than one strong particle in one minimum (Figs.~\ref{fig:3}d). This allows the spatial rearrangement of the particles since the energy of the interaction with the speckle is no longer strong enough to localize the particles, a large scale crystalline structure is favorable, consistent with the increase in $\langle\phi_6\rangle$ observed in Fig.~\ref{fig:2}. 
When $\chi=1$, all particles are strong and thus the hexagonal crystalline structure is recovered.

\begin{figure*}[ht]
\includegraphics[width=\textwidth]{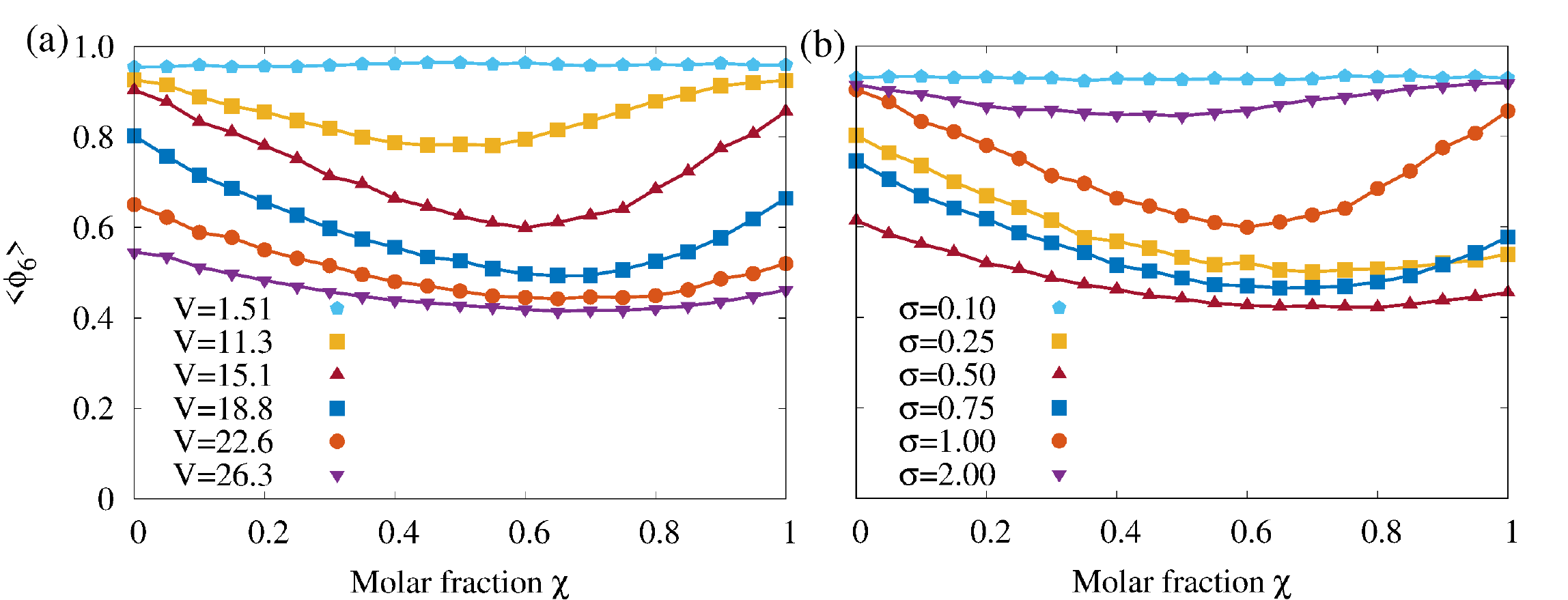}
\caption{\textbf{Dependence of the order parameter on the speckle properties.}
Six-fold bond order parameter as a function of the molar fraction ($\chi$) obtained numerically, for different values of the speckle (a) strength and (b) spatial correlation $\sigma$. Results in (a) were obtained for $\sigma=1$ and in (b) for $V=15.1$, and are averages over $100$ samples.
}
\label{fig:4}
\end{figure*}

In order to explore how robust this effect is, we studied numerically how it depends on the properties of the underlying speckle pattern. The speckle is characterized by a strength $V$ corresponding to the average potential depth (in units of $k_{\rm B} T$, where $k_{\rm B}$ is the Boltzmann constant and $T$ is the absolute temperature of the sample) and by a spatial correlation $\sigma$ (in units of the particle diameter), which corresponds to the average grain size. 
Figure~\ref{fig:4}(a) shows $\langle\phi_6\rangle$ for different $V$. Although the curves in the range $1.51<V\leqslant18.8$ feature one minimum, its position and intensity vary with $V$: when $V$ increases, the number of strong particles that can be trapped increases monotonically and, consequently, $\chi_\mathrm{min}$ shifts to the right and the minimum becomes deeper. For $V>18.8$, the behaviour seems to become independent of the molar fraction (and always disordered), because the weak particles are also strongly trapped.
Figure~\ref{fig:4}(b) shows $\langle\phi_6\rangle$ for different values of $\sigma$. A pronounced minimum is only observed for intermediate values of $\sigma$, close to unity (particle diameter). If $\sigma \gg 1$ or $\sigma \ll 1$, the optical forces are negligible for different reasons: for $\sigma \gg 1$, the gradient of the optical potential is very small on the scale of the particle; and for $\sigma \ll1$, the optical potential varies on a length scale smaller than the particle size and thus its gradient averages to zero over the particle cross-section (see supplementary Fig.~S3 \cite{SM}). In the latter case, the optical force on a particle is the sum of the contributions over the particle's cross-section, which can be described by an effective random potential that differs from the one originally applied (Supplemental Material and supplementary Fig.~S4 and S5 \cite{SM}). 

In conclusion, we have shown that the order in a two-dimensional binary colloidal crystal can be controlled by an underlying random optical potential, when each species experiences distinct optical forces. 
Since the intensity of the optical forces depends on the mismatch of the indices of refraction of the particles and the surrounding medium, the particles with the larger index mismatch are more responsive (strong particles) than those with the lower mismatch (weak particles). For the parameters of the optical potential that were considered, only the strong particles respond significantly to the potential. Thus, strong particles tend to occupy the minima of the potential and nucleate structural defects in the, otherwise, periodic hexagonal structure of the weak particles. The density of defects is controlled by the fraction of strong particles and the statistical properties of the underlying potential. When the number of strong particles increases beyond the number of local minima that can trap them, the trapping mechanism becomes less effective and the hexagonal order is recovered as the fraction of strong particles increases. Here, we have considered a random optical potential with Gaussian spatial correlations and with a characteristic length given by the standard deviation $\sigma$. However, it is technically possible to generate other optical potentials, e.g. periodic~\cite{Jones2015} or with different spatial correlations~\cite{Bromberg2014, Bender2018}. Thus, one can control not only the density of defects, but also their spatial distribution. Time-varying optical potentials could also be employed to change the position of strong particles and defects in time, affecting the overall dynamics \cite{Grier2003, Volpe2014OE, Brazda2017}. Understanding how the spatial distribution of defects influences the physical properties of materials is a question of both scientific curiosity and technological interest that can now be addressed in a systematic way.

\begin{acknowledgments}
Andr\'e S. Nunes, Margarida M. Telo da Gama and Nuno A. M. Ara\'ujo acknowledge financial support from the Portuguese Foundation for Science and Technology (FCT) under Contracts nos. EXCL/FIS-NAN/0083/2012, UID/FIS/00618/2019, SFRH/BD/119240/2016 and PTDC/FIS-MAC/28146/2017 (LISBOA-01-0145-FEDER-028146). Margarida Telo da Gama and Nuno Araújo would like to thank the Isaac Newton Institute for Mathematical Sciences for support and hospitality during the program "The mathematical design of new materials" where the final version of this manuscript was completed. This program was supported by EPSRC Grant Number: EP/R014604/1. Giorgio Volpe acknowledges support from the Royal Society under grant RG150514. We also acknowledge Parviz Elahi for his help with the experimental setup.
\end{acknowledgments}

\bibliography{refs}

\end{document}